\documentclass[12pt]{article}
\usepackage{fullpage}
\newcommand\dw{\dot{w}}
\newcommand\pr{\prime}
\newcommand\be{\begin{equation}}
\newcommand\ee{\end{equation}}
\newcommand\bea{\begin{eqnarray}}
\newcommand\eea{\end{eqnarray}}
\newcommand\nn{\nonumber}

\newcommand\bdm{\begin{displaymath}}
\newcommand\edm{\end{displaymath}}
\newcommand\dts{\mbox{\raisebox{1.75ex}{
 {\mbox{\raisebox{.25ex}{$\cdot$}\raisebox{-1ex}{$\!\widetilde{}$}}}}}{\!\!\!S}}

\def\pmb#1{\setbox0=\hbox{#1}%
  \kern-.025em\copy0\kern-\wd0
   \kern.05em\copy0\kern-\wd0
   \kern-.025em\raise.0433em\box0 }
\def\bx{\mbox{\boldmath$x$}}
\def\by{\mbox{\boldmath$y$}}

\def\bw{\mbox{\boldmath$w$}}
\def\bv{\mbox{\boldmath$v$}}
\def\bVV{\mbox{\boldmath$V$}}
\def\bb{\mbox{\boldmath$b$}}
\newcommand\bnab{\mbox{\boldmath$\nabla$}}

\begin{document}

\title{\large {\bf  A DYNAMICAL THEORY OF MARKOVIAN DIFFUSION\thanks{{\em Physica\/} {\bf 96A}  (1979) 465--489 \copyright {\em North-Holland Publishing Co.}}}}

\author{Mark DAVIDSON\thanks{
Current Address: Spectel Research Corporation, 807 Rorke Way, Palo Alto, CA   94303
\newline  Email:  mdavid@spectelresearch.com, Web: www.spectelresearch.com}\\
\normalsize{\em  }}

\date{ \normalsize Received 29 March 1978\\Revised 23 October 1978 }

\maketitle

{\small A dynamical treatment of Markovian diffusion is presented and several applications discussed.  The stochastic interpretation of quantum mechanics is considered within this framework.  A model for Brownian movement which includes second order quantum effects is derived.}

\section{Introduction}

Continuous Markov processes, a well developed field of mathematics \cite{doob}--\cite{levy},  have not yet had their full impact on modern physics.  The earliest of these, developed by Einstein \cite{einstein}, Smoluchowski \cite{smoluchowski}, and later refined by Wiener \cite{wiener} were the original attempts to model diffusion.  They did not have a simple phase space interpretation, and as a result were considered unphysical.  The Ornstein-Uhlenbeck model \cite{orn-uhl} which did allow a phase space picture, but was not a Markov process in coordinate space, became the preferred description.  Although mathematicians pursued the field vigorously, physicists consequently lost interest.  The work of Nelson \cite{nelson1,nelson2} was a noteworthy exception.  He studied the general problem of diffusion and found in a particular model that if a quantity called the mean acceleration were equated with the external force, paralleling Newton's equation, then Schr\"odinger's equation could be derived, suggesting a connection with quantum theory.  The results presented here are consistent with this, although a different approach is taken to the  same model, and a different dynamical assumption is made.

In section 2 a new technique of analysis is developed for the simple but important case of the Wiener process.  Velocities as random variables cannot be defined for these processes.  The sample trajectories of the diffusing particles are nowhere differentiable almost surely.  A Hilbert space approach is presented in which coordinates, velocities, and accelerations are noncommuting operators.  This is an improvement over an earlier work which introduced the concept of ordered expectation \cite{davidson}, and which presented many of the basic ideas of the Hilbert space method.  The various operators have a physical interpretation as they are defined in terms of limits of expectations of random variables.  This method is used to derive several known results to illustrate consistency.

In section 3 the Hilbert space method is applied to the general Markov diffusion model.  Operators for coordinates, velocities, and accelerations are defined in the same way as for the Wiener process, and they are related to the properties of the underlying stochastic model.  Commutation rules are derived and equations governing the time evolution of the diffusion presented.  Section 4 generalizes the technique to three dimensions.  The approach is easily generalized to any system with a finite number of degrees of freedom.

The dynamical assumption, introduced in section 5, equates the acceleration operator to a sum of external forces plus a stochastic force.  This stochastic force is chosen so as not to violate momentum and angular momentum conservation in the absence of an external force and in the absence of viscosity.  The resulting model has four parameters.  By varying these, models are obtained for Schr\"odinger's equation and for thermal diffusion.

It is found that Schr\"odinger's equation cannot be identified with a unique Markov process, but rather an infinite number of different processes can lead to it.  This curious fact suggests that the stochastic interpretation of quantum mechanics be modified to incorporate this non-uniqueness and perhaps resolve some of the conceptual difficulties which the interpretation has faced in the past.  The non-commutative approach developed here, so similar to quantum mechanics, increases the need to explore completely the connection between these two theories.

The thermal model approaches the quantum mechanical Gibbs distribution in the steady state limit, with error terms of order  $\hbar^4$, as it should.  It can be generalized to include viscosity, and it has a Newtonian limit when diffusive fluctuations are small.  The usual diffusion approximation is found to be a good one under certain circumstances.

A method for solving the diffusion equations is presented which aids in understanding and which utilizes a mathematical similarity between the diffusion theory and the Hamilton-Jacobi approach to classical mechanics.

\section{The Wiener process}

A collection of measurable functions $w(t)$, $t\in (-\infty, \infty)$ on a probability space  $(\Omega, \Sigma, P)$ is called a Wiener process with   zero drift if
\bea
&&P(w(t) -w(s)  \;|\; w(r)) = P(w(t) -w(s) ) , \quad r<s, t, \label{1}\\
&&E(w(t)) =0, \quad \mbox{ all } t.
\label{2}
\eea
When specialized to the case $w(0)=0$ identically it becomes a  Gaussian Markov process for $t>0$:
\bea
&&\rho(w, t)=\frac{1}{\sqrt{2\pi v t}} \; e^{-w^2/2\nu t}, \quad \nu=\mbox{ constant},\label{3}\\
&&E(w(t_1) w(t_2)) =\nu \min(t_1, t_2), \label{4}\\
&&E(w(t_1) \times \cdots \times w(t_n))\nn\\
&&\qquad\qquad\qquad =\frac{1}{2^{n/2}} \; \frac{1}{(n/2)!} \; \sum\limits_\pi \; E(w(t_1) w(t_2) ) \times \cdots \times E (w(t_{n-1}) w(t_n)),
\label{5}
\eea
where $\pi$ denotes a permutation of the $t$ subscripts and (\ref{5}) vanishes for $n$ odd.  The Markov property is
\be
P(w(t) \in B \; | \;w(s_1) , \ldots, w(s_n)) =P(w(t) \in B \; | \;w(s_1)), \quad i=1, n,
\label{6}
\ee
where $B$ is an arbitrary Borel set, and $t>s_i >s_{i+1}$, all  $i$.

The density $\rho$ satisfies the Fokker-Planck or continuity equation
\be
\frac{\partial}{\partial t} \rho =\frac{1}{2} \nu \frac{\partial^2}{\partial x^2}\rho.
\label{7}
\ee
The Markov transition function for this process, expressed as a density, is
\be
P_{t-s} (x,y)=\frac{1}{\sqrt{2\pi \nu (t-s)}}\; \exp \left( -\frac{(x-y)^2}{2\nu (t-s)}\right), \quad t>s,
\label{8}
\ee
which satisfies the Chapman-Kolmogorov equation
\be
P_s (x, y) =\int \,{\rm d}z\, P_{s-u} (x, z) P_u (z,y).
\label{9}
\ee
Joint probability densities may be constructed with the transition functions.  For example, the joint density for $w(t)=y$ and $w(s) =x$, $\, t>s$ is
\be
\rho (x, s; y,t) =\rho (x, s) P_{t-s} (x,y).
\label{10}
\ee

It is well known that the sample trajectories for $w$ are nowhere differentiable a.s.  As a consequence, instantaneous velocity does not exist as a random variable.  Therefore define
\be
\dot{w}(t)=\frac{w(t+ \epsilon /2) -w(t-\epsilon /2)}{\epsilon},
\label{11}
\ee
which is a random variable for $\epsilon >0$, but loses this property in the limit $\epsilon \to 0$.  Using eq. (\ref{4}) one finds
\bea
&&E(\dot{w}(t) w(t+\delta)) =\nu, \quad \delta> \epsilon /2,\label{12}\\
&&E(\dot{w} (t) w(t-\delta)) =0, \quad \delta>\epsilon /2.
\label{13}
\eea
These equations remain true in the limit $\epsilon$, $\delta\to 0$ provided the inequalities are respected.  They suggest that a non-commutative approach may be a possible way to study the  Wiener process.

Let $H_t$ be the completed real Hilbert space of functions $f(x)$ with inner  product
\be
(f,g) =E (f(w(t))g(w(t)))=\int \; {\rm d}x \, \rho (x,t) f(x) g(x).
\label{14}
\ee
$H_t$ is spanned by the Hermite polynomials.  Define a linear operator $\dot{w}$ on $H_t$ by
\be
(f, \dot{w} g)= \lim\limits_{\stackrel{\epsilon, \delta_i \to 0_+}{\epsilon / \delta_i \to 0}} \; E(f(w(t-\delta_1))\dot{w} (t) g(w(t+\delta_2))),
\label{15}
\ee
where $f, g\in H_t$.  This definition is independent of how the $\delta_i$ are taken to zero because the expectation in (\ref{15}) satisfies
\bea
&& E(f\dot{w} g) = \int \; {\rm d}x\, {\rm d}y\, \rho (x,t -\delta_1; y, t +\delta_2) f(x) g(y)\nn\\
&&\quad \times \; E (\dot{w}(t) \; |\; w(t-\delta_1) = x, \quad w(t+\delta_2)=y).
\label{16}
\eea
This equation follows from the Markov property together with the definition of conditional expectation.  It is easy to show
\be
E(\dw (t)\; |\; w(t-\delta_1) =x, w (t +\delta_2) = y ) = \frac{y-x}{\delta_1+ \delta_2},
\label{17}
\ee
and so
\bea
&& (f, \dw g) =\lim \int \, {\rm d}x\, {\rm d}y\, \exp \left( -\frac{x^2}{2\nu (t-\delta_1)}\right) \exp \left( -\frac{(y-x)^2}{2\nu (\delta_1 +\delta_2)}\right) \; \frac{y-x}{\delta_1 +\delta_2}\nn\\
&&\qquad \times f(x) g(y) \frac{1}{\sqrt{2\pi\nu}} \,\left[\left(t-\delta_1\right) \left( t+\delta_2\right)\right]^{-1/2}.
\label{18}
\eea
In this expression $t-\delta_1$ may be replaced by $t$ (for $t>0$) without affecting the limit and the resulting expression is a function of $\delta_1 +\delta_2$ so that the manner in  which the limit is taken doesn't matter.

Suppose that $g(y)$ is a polynomial.  It has a Taylor's expansion about $x$
\be
g(y) = g(x) +(y-x) g^\pr (x) + \cdots.
\label{19}
\ee
When substituting this expansion into (\ref{18}), only the second term contributes.  One finds
\be
(f, \dw g)= \left( f, \nu \frac{\partial g}{\partial x}\right) .
\label{20}
\ee
Since $f$ is any element of $H_t$, if follows that
\be
\dw g=\nu \frac{\partial g}{\partial x}, \quad g \mbox{ a polynomial}.
\label{21}
\ee
If an operator $w$ is defined by simple multiplication, then it follows
\be
\left[ \dw, w\right] g=\nu g , \quad g \mbox{ a polynomial}.
\label{22}
\ee

Similarly, an operator for acceleration may be defined.  First define a random variable
\be
\ddot{w} (t) =\frac{w (t+\epsilon)+ w(t-\epsilon) -2w(t)}{\epsilon^2},
\label{23}
\ee
and next define an operator on $H_t$
\be
(f, \ddot{w}g)=\lim\limits_{\stackrel{\epsilon, \delta_i\to 0_+}{\epsilon/\delta_i\to 0}}\; E (f(w(t-\delta_1)) \ddot{w} (t) g(w(t+\delta_2))).
\label{24}
\ee
It is easy to show that this vanishes from the fact that
\bdm
E (\ddot{w} (t) \;| \; w (t-\delta_1) =x , w(t+\delta_2) =y)=0,
\edm
which follows from eq. (\ref{17}).  Therefore
\be
\ddot{w}=0,
\label{25}
\ee
as an operator equation.  Likewise all operators for higher time derivatives defined in this way are zero.

Let us define
\be
H=\frac{1}{2} \dw^2.
\label{26}
\ee
$H$ is not symmetric when restricted to the polynomials
\be
(f, Hg) =(Hf\rho , g/\rho),
\label{27}
\ee
and it satisfies
\be
[H, w] = \nu \dw.
\label{28}
\ee
$H$ governs the time evolution of expectations as can be seen by theorems I and II.

\noindent {\em Theorem I\/}   For $f$ a polynomial
\be
\frac{\rm d}{{\rm d}t} \, E(f(w(t)))=\left( 1, \frac{1}{\nu} \left[ H, f(w)\right]\right).
\label{29}
\ee

\noindent {\em Proof\/}
\bea
\left( 1, \frac{1}{\nu} \left[ H, f(w)\right] \right) &=& \int\, {\rm d}x\, \rho (x,t) \frac{\nu}{2} \frac{\partial^2}{ \partial x^2} \, f(x)\label{30}\\
&=& \int \, {\rm d}x\, \frac{1}{2} \nu f (x) \frac{\partial^2}{\partial x^2} \, \rho (x,t)\label{31}\\
&=& \int \, {\rm d}x\, f(x) \frac{\partial}{\partial t} \, \rho (x,t) =\frac{\rm d}{{\rm d}t} \, E(f(w(t))),
\label{32}
\eea
where the Fokker-Planck equation (\ref{7}) has been used in going from (\ref{31}) to (\ref{32}).

The following operators are well defined when acting on polynomials.
\be
w(t+s) =e^{(1/\nu)Hs} w \, e^{-(1/\nu)Hs} =w + s\dw.
\label{33}
\ee
They satisfy commutation rules
\be
\left[ w\left(u_1\right) , w\left( u_2\right)\right] =\nu (u_1 -u_2).
\label{34}
\ee
These operators play a role similar to the coordinate operators in the Heisenberg representation of quantum mechanics as can be seen from theorem II.

\noindent{\em Theorem II\/}
\be
E(w(t_1) \times \cdots \times w(t_n)) = (1, w(t_1) \times \cdots \times w(t_n)), \quad t_i < t_{i+1}.
\label{35}
\ee

\noindent {\em Proof\/} Let $s_i=t_i -t$.  Using (\ref{33}) one has
\bdm
(1, w(t_1) \times \cdots \times w(t_n)) =\int \, {\rm d}x \, \rho(x,t) \left( x + s, \nu \frac{\partial}{\partial x}\right) \times \cdots \times \left( x + s_n \nu \frac{\partial}{\partial x}\right).
\edm
Integration by parts of the right hand side, using the Fokker-Planck equation, yields
\bdm
\int \, {\rm d}x \, \rho (x, t) (t+s_1) \nu \, \frac{\partial}{\partial x} \, \left( x + s_2\nu \frac{\partial}{\partial x} \right) \times \cdots \times \left( x+ s_n \nu \frac{\partial}{\partial x}\right).
\edm
But, using (\ref{4}), this may be written
\be
(1, w(t_1) \times \cdots \times w(t_n))=\sum\limits^n_{i=2} \, E(w(t_1) w(t_i)) \left(1, T\prod\limits^n_{\stackrel{j=2}{j\not= i}} w(t_j)\right).
\label{36}
\ee
where $T$, a notational device, denotes time ordering of the product, with later times standing to the right.  From eq. (\ref{5}) straightforward combinatorics yields
\be
E(w(t_1) \times \cdots \times w(t_n) ) =\sum\limits^n_{i=2} \, E (w(t_1) w(t_i)) E \left( \prod\limits^n_{\stackrel{j=2}{j\not= i}} w(t_i)\right).
\label{37}
\ee
Subtracting (\ref{37}) from (\ref{36}) yields
\bdm
D_n (t_1, \ldots, t_n) =\sum\limits^n_{i=2} E(w (t_1) w(t_i)) D_{n-2} \left( \left\{ t_j \, s.t.\,  j\not= 1, j\not= i\right\} \right).
\edm
where
\bdm
D_n (t_1, \ldots, t_n) =\left( 1, T \prod\limits^n_{j=1} w(t_j)\right) -E(w (t_1) \times \cdots \times w(t_n)).
\edm
But, $D_1=D_2=0$ by direct computation, and the theorem follows immediately by transfinite induction.

As a corollary to theorem II, for $f_i$ polynomials:
\bdm
E(f_1(w(t_1))\times \cdots \times f_n (w(t_n))) =(1, f_1(w(t_1)) \times \cdots \times f_n (w (t_n))), \quad t_{i+1} > t_i.
\edm
which follows by expanding the polynomial products using theorem II.

For the Wiener process, time translations correspond to similarity transformations on the coordinates as in (\ref{33}). It is interesting to compare this with quantum mechanics where time translations correspond to unitary transformations.  This is a fundamental difference.  For example, a product of different time coordinates in quantum mechanics is not hermitian and does not even have a real expectation.

An ordered expectation may be defined in the following way.  Let
\bdm
w_1 (t) =w(t), \quad w_2(t) =\dw(t), \quad w_3 (t) =\ddot{w} (t)
\edm
and define
\be
\overline{w_{j_1}(t) \times \cdots \times w_{j_n} (t)} =\lim\limits_{\stackrel{t_i\to t, t_{i+1}>t_i}{\epsilon/(t_{i+1}-t_i)\to 0_+}} \; E(w_{j_1} (t_1) \times \cdots \times w_{j_n} (t_n)).
\label{38}
\ee
Ordered polynomials $\overline{f(w, \dw)}$ are therefore defined.  Using theorem II it follows
\be
\overline{f(w, \dw)} =(1, f(w, \dw)) = \int {\rm d}x\, \rho (x, t) f\left( x, \nu \frac{\partial}{\partial x}
\right).
\label{39}
\ee
Likewise, for ordered polynomials involving $w$, $\dw$, and $\ddot{w}$ one finds
\be
\overline{f(w, \dw ,\ddot{w})} =(1, f(w, \dw, \ddot{w}))=\int\, {\rm d}x\, \rho (x,t) f \left( x, \nu \frac{\partial}{\partial x}, 0\right).
\label{40}
\ee

These ordered expectations satisfy a kind of completeness relation.  Defining
\bea
\overline{f(w, \dw) \delta (w-z)} &=& \int \,{\rm d} x\, \rho (x,t) f\left( x, \nu \frac{\partial }{\partial x} \right)\delta (x-z),\label{41}\\
\overline{\delta(w-z) f(w,\dw)} &=& \int\, {\rm d}x\, \rho (x,t) \delta (x-z) f\left( x, \nu \frac{\partial}{\partial x}\right),
\label{42}
\eea
the following results are obtained after several integrations by parts:
\be
\int \, {\rm d}z\, \frac{1}{\rho(z,t)}\, \overline{f(w, \dw) \delta (w-z)} \; \overline{\delta(w-z) g (w, \dw)} =\overline{f(w, \dw)g(w, \dw)}.\label{43}
\ee
This result is similar to quantum mechanics where states of definite coordinate form a complete set of states for a spinless non-relativistic particle.

The $\delta$ function can be used, at least formally, to derive the correct probability densities.  For example,
\bea
(1, \delta(w(t_1) -y)) &=& \int \, {\rm d}x\, \rho (x,t) \delta (e^{(1/\nu) H(t_1-t)} x \, e^{-(1/\nu)H(t_1-t)} -y)\nn\\
&=& \int \, {\rm d}x\, \rho (x,t) e^{(1/\nu) H (t_1-t)} \delta (x-y) \,e^{-(1/\nu) H(t_1 -t)}\nn\\
&=& \int\, {\rm d}x \, \delta (x-y) e^{(1/\nu) H(t_1 -t)} \rho (x,t)\nn\\
&=& \rho (y, t_1), \quad H=\frac{1}{2} \nu \,\frac{\partial^2}{\partial x^2}.
\label{44}
\eea
Likewise,
\be
(1, T\delta(w(t_1) -y) \delta (w(t_2)-z))=\rho (y, t_1; z, t_2),
\label{45}
\ee
and generally
\be
(1, T\delta (w, t_1) -y_1 ) \times \cdots \times \delta (w (t_n) - y_n))=\rho(y_1, t_1 ; \ldots; y_n, t_n).
\label{46}
\ee

\section{Generalized Brownian motion}

The non-commutative approach outlined for the Wiener process has an analog for a deeper class of stochastic processes defined formally by the stochastic differential equation:
\be
{\rm d}x(t) =b(x(t), t) {\rm d}t+ {\rm d}w(t).
\label{47}
\ee
It is more precise to consider a stochastic integral equation which is formally equivalent to this:
\be
x(t) -x(0) =\int\limits^t_0\, b(x(t^\pr), t^\pr) {\rm d}t^\pr +w(t).
\label{48}
\ee
This equation has been studied extensively in the literature \cite{doob}--\cite{levy}.  Nelson \cite{nelson1,nelson2} examined the relationship between this theory and non-relativistic quantum mechanics.

Higher dimensional diffusion will be considered, but first the one dimensional case shall be examined.  Doob \cite{doob} has shown that (\ref{48}) is solved by iteration provided $b(x,t)$ satisfies certain conditions:  it must be a Baire function in the variables $x$ and $t$, it must satisfy the bound $b\leq k\sqrt{1+x^2}$, and it must satisfy a uniformity condition:
\bdm
|b(x_1, t)-b(x_2, t)|\leq K |x_2-x_1|
\edm
for some constant $K$. He also showed that $x(t)$ is a Markov process and that its sample trajectories are almost all continuous.

The Markov transition function plays an important role in the analysis of this process.  Rather than working with the usual transition function, which is a distribution, it is convenient to work with its associated density, assuming that this exists. The transition density satisfies a Chapman-Kolmogorov equation:
\be
P(x,t; y, s) =\int\, {\rm d}z\, P (x,t: z, u) P (z, u; y, s), \quad t>u>s
\label{49}
\ee
where
\be
P(x,t; y,s) =\frac{\partial}{\partial x} \,P(x(t) <x \,| \,x(s) =y), \quad t>s.
\label{50}
\ee

With regularity assumptions, Doob derives a forward (FE) and backward (BE) equation for $P$:
\bea
\mbox{FE}: &&\frac{\partial}{\partial t}\, P (x,t; y,s) +\frac{\partial}{\partial x} \, b (x,t) P (x, t; y,s) -\frac{1}{2} \nu \frac{\partial^2}{\partial x^2} \, P (x,t; y,s) =0, \label{51}\\
\mbox{BE}: && \frac{\partial}{\partial s}\, P(x,t: y, s) +b(y,t) \,\frac{\partial}{\partial y} \, P(x,t; y,s) +\, \frac{\nu}{2} \,\frac{\partial^2}{\partial y^2} \, P(x,t; y,s) =0,
\label{52}
\eea
where $t>s$, and $\nu$ is the diffusion parameter for $w$ in eqs. (\ref{47}) and (\ref{48}).  He also shows
\be
\lim\limits_{h\to 0_+} \, E\left( \frac{x(t+h) -x(t)}{h}\biggl|\biggr. x(t) =x\right) =b(x,t)
\label{53}
\ee
and
\be
\lim\limits_{h\to 0_+} \,  E\left( \frac{(x(t+h) -x(t))^2}{h} \biggl|\biggr. x(t) =x\right) =\nu.
\label{54}
\ee

With these results in mind, let $\rho (x,t)$ denote the probability density for $x(t)$ at time $t$ (assuming that it exists)
\be
\rho(x,t) =\frac{\partial}{\partial x} \, P (x(t) \leq x).
\label{55}
\ee
Denote by $H_t$ the real Hilbert space of functions on $R$ with inner product
\be
(f,g) =\int\limits_R \, {\rm d}x \,f(x) g(x) \rho (x,t) =E(f(x(t) ) g(x(t))).
\label{56}
\ee
It shall be assumed that $\rho(x,t)$ is a function of fast decrease as $x \to\infty$.  Therefore, $H_t$ contains all of the polynomials, and these are dense in $H_t$.  Define random variables:
\bea
\dot{x} (t) &=&\frac{x(t+\epsilon/2)-x(t-\epsilon /2)}{\epsilon},\label{57}\\
\ddot{x} (t) &=& \frac{x(t+\epsilon) +x(t-\epsilon) -2x(t)}{\epsilon^2}.
\label{58}
\eea
Define an operator $\dot x$ on $H_t$ which has matrix elements for $f$ and $g$  polynomials:
\be
(f, \dot{x} g) =\lim\limits_{\stackrel{\stackrel{\epsilon\to 0_+}{\delta_1, \delta_2\to 0_+}}{\epsilon /\delta_i\to 0}} \, E(f(x(t-\delta_1) \dot{x} (t) g (x(t+\delta_2))).
\label{59}
\ee
Using the Markov property, which is the same as (\ref{6}), it follows
\bea
&& E(f(x(t-\delta_1))\dot{x} (t) g(x(t+\delta_2)))\nn\\
&&\quad = \int \, {\rm d}x \, {\rm d}y\, \rho (x, t -\delta_1) P (y, t+\delta_2 ; \, x, t -\delta_1) f(x) g(y)\nn\\
&&\qquad \times E (\dot{x} (t)\, |\, x(t-\delta_1) = x, x(t+\delta_2) =y).
\label{60}
\eea
Using the forward and backward equations, the conditional expectation in (\ref{60}) can be calculated.  The result is that (\ref{60}) can be rewritten as
\bea
&=& \int\, {\rm d}x\, {\rm d}y \, {\rm d}z\, \rho (x,t-\delta_1) f(x) g(y) \biggl\{ P(z, t; x, t-\delta_1)\biggr.\nn\\
&&\quad \times  P(y, t+\delta_2; z, t) b (z,t) +P (z,t; x,t -\delta_1)\nu \,\frac{\partial}{\partial z} \, P (y, t +\delta_2 ; z,t)\biggl.\biggr\},
\label{61}
\eea
where the $\epsilon\to 0$ limit has already been taken.  Since the sample paths are continuous, it follows that
\be
\lim\limits_{t\to s} P(x,t; y,s) =\delta (x-y).
\label{62}
\ee
For the limit $\delta_1, \delta_2\to 0$, (\ref{62}) can be substituted into (\ref{61}).  The result is
\be
(f, \dot{x} g)= \int \, {\rm d}x\, \rho (x,t) f(x) \left(b(x,t) +\nu\,\frac{\partial}{\partial x}\right) \, g(x)
\label{63}
\ee
which may be rewritten
\be
(f, \dot{x} g) =\int \; {\rm d}x\, e^{R-S} f(x) \nu \, \frac{\partial}{\partial x} g(x) e^{R+S},
\label{64}
\ee
where $R$ and $S$ are defined by
\bea
R&=& \frac{1}{2} \ln (\rho (x,t)),\label{65}\\
b&=& \nu \frac{\partial}{\partial x} \left(R+S\right)
\label{66}
\eea
and $S$ is defined only up to an arbitrary additive function of $t$.

An operator for acceleration is similarly defined as follows
\be
(f, \ddot{x} g ) =\lim\limits_{\stackrel{\epsilon, \delta_1, \delta_2\to 0_+}{\epsilon / \delta_i\to 0}}\, E(f(x(t-\delta_1))\ddot{x} (t) g(x (t+\delta_2)))
\label{67}
\ee
but,
\bea
&& E (f(x(t-\delta_1))\ddot{x} (t) g(x(t+\delta_2)))=\int \, {\rm d}x\, {\rm d}y \, \rho (x,t-\delta_1)\nn\\
&&\quad \times P(y,t +\delta_2; x, t -\delta_1)f (x) g(y) E(\ddot{x} (t) \, | \, x(t-\delta_1) =x, x(t+\delta_2) =y).
\label{68}
\eea
Using the forward and backward equations and the definition of conditional expectation, this can be rewritten as
\bea
&&\int \, {\rm d}x\, {\rm d}y \, {\rm d}z\, \rho (x,t-\delta_1) f(x) g(y) P  (z, t; x, t-\delta_1) P (y, t+\delta_2; z, t)\nn\\
&&\quad \times \left[ \frac{\partial b (z, t)}{\partial t} + \frac{1}{2} \nu \,\frac{\partial^2 b(z,t)}{\partial z^2} +\frac{1}{2} \, \frac{\partial}{\partial z} \, b^2 (z,t) \right].
\label{69}
\eea
Substituting the $\delta$ function of (\ref{62}) into this expression yields the proper limit for (\ref{67}).  One finds
\be
(f, \ddot{x} g)= \int \, {\rm d}x\, \rho (x,t) f(x) g(x) \left[\frac{\partial b(x,t)}{\partial t} +\frac{1}{2} \nu \, \frac{\partial^2 b(x,t)}{\partial x^2} +\frac{1}{2} \, \frac{\partial}{\partial x} b^2\right].
\label{70}
\ee
So that
\be
\ddot{x} =\frac{\partial b}{\partial t} +\frac{1}{2} \nu \,\frac{\partial^2b}{\partial x^2} +\frac{1}{2} \, \frac{\partial}{\partial x} b^2,
\label{71}
\ee
or defining a stochastic potential $U$ by
\be
-\frac{\partial U}{\partial x} =\ddot{x}
\label{72}
\ee
(\ref{71}) becomes
\be
\frac{\partial U}{\partial x} + \frac{\partial b}{\partial t} +\frac{1}{2} \nu \, \frac{\partial^2b}{\partial x^2} +\frac{1}{2} \frac{\partial}{\partial x} b^2=0.
\label{73}
\ee
But using (\ref{66}) this can be transformed into:
\be
\left[\frac{1}{2} \nu^2 \, \frac{\partial^2}{\partial x^2} +U(x,t) \right] e^{R+S} =-\nu \frac{\partial}{\partial t} \, e^{R+S}.
\label{74}
\ee
To supplement this equation, write the density at time $t$ as
\be
\rho(x,t) =\int \; {\rm d}y \, \rho (y,s) P (x,t; y,s) , \quad t>s.
\label{75}
\ee
Therefore, $\rho$ must satisfy the forward equation
\be
\frac{\partial \rho}{\partial t} +\frac{\partial}{\partial x} b\rho -\frac{1}{2} \nu \, \frac{\partial^2}{\partial x^2} \,\rho =0.
\label{76}
\ee
Using (\ref{65}) and (\ref{66}) this becomes
\be
\frac{\partial \rho}{\partial t} =-\frac{\partial}{\partial x} \,\nu\,\frac{\partial S}{\partial x}\, \rho.
\label{77}
\ee
Equations (\ref{74}) and (\ref{77}) are easily shown to be equivalent to
\be
\left[ \frac{1}{2}\,\nu^2 \, \frac{\partial^2}{\partial x^2} +U\right] \, e^{R\pm S} =\mp \nu\,\frac{\partial}{\partial t} \, e^{R\pm S}.\label{78}
\ee
These are the basic form of the diffusion equations which will be used here.  They are referred to as the Markov wave equations.  They may be understood heuristically in the following way.  Ordered expectations may be defined for $x(t)$, $\dot{x}(t)$, and $\ddot{x}(t)$ as in eq.\ (\ref{38}).  For $f$ an ordered polynomial, one finds
\bea
\overline{f(x, \dot{x} , \ddot{x})} &=& \left( 1, f \left( x, \dot{x} , -\frac{\partial U}{\partial x}\right)\right)\nn\\
&=& \int \, {\rm d}x\, e^{R-S} f\left( x, \nu \frac{\partial}{\partial x}, -\frac{\partial U}{\partial x} \right) e^{R+S}.
\label{79}
\eea
The commutation rules implicit in (\ref{63}) and (\ref{64}) are
\be
\left[\dot{x}, x\right] =\nu,
\label{80}
\ee
and from these one deduces the following rule, using the chain rule of differentiation together with the definition of ordered expectations:
\be
\frac{\rm d}{{\rm d}t} \,\overline{f(x, \dot{x})} =\frac{\rm d}{{\rm d}t} (1, f(x, \dot{x} )) =(1, [H, f(x, \dot{x} )])/\nu,
\label{81}
\ee
where
\be
H=\frac{1}{2} \dot{x}^2+U.
\label{82}
\ee
But (\ref{81}) is equivalent to
\bea
&&\frac{\rm d}{{\rm d}t} \int\, {\rm d}x \, e^{R-S}\, f\, \left( x, \nu \frac{\partial}{\partial x} \right) \,e^{R+S}\nn\\
&&\,\quad =\int \, {\rm d}x \, e^{R-S} \left[ \frac{1}{2} \nu \, \frac{\partial^2}{\partial x^2} +U, f\left( x, \nu\, \frac{\partial}{\partial x}\right)\right] e^{R+S}
\label{83}
\eea
which must be satisfied for arbitrary $f$.  This is possible only if the Markov wave equations (\ref{78}) are satisfied.

The situation is similar to quantum mechanics. The arena of the theory is a Hilbert space $H_t$.  Coordinates and velocities are operators on this space with commutation rules given by (\ref{80}).  There is an operator $H$ which determines time derivatives.

The Wiener process is a special case of this theory.  When $b=0$ then it is seen from (\ref{48}) that $x(t)$ is a Wiener process, provided $x(0)$ is identically zero.  If $b$ is zero then one may chose $R+S=0$.  The Markov wave equations become
\be\
\left( \frac{1}{2} \nu^2 \,\frac{\partial^2}{\partial x^2} +U\right) e^{R\mp R} =\mp \nu \frac{\partial}{\partial t} e^{R\mp R},
\label{84}
\ee
which force that $U=0$, and
\be
\frac{1}{2} \,\nu \, \frac{\partial^2}{\partial x^2} \, e^{2R} =\frac{\partial}{\partial t} \, e^{2R},
\label{85}
\ee
which is just the diffusion equation (\ref{7}).  Eq.\ (\ref{64}) becomes
\be
(f, \dot{x} g)=\int \; {\rm d}x\, e^{2R} f(x) \nu \, \frac{\partial }{\partial x} g(x) ,
\label{86}
\ee
which agrees with      eq.\ (\ref{20}).  If $b=0$, then $\ddot{x} =0$ from (\ref{71}), which is consistent with the Wiener result $\ddot{w} =0$ [(25)].  Thus the general theory consistently describes the Wiener process as it must.

\section{Three dimensions}

In three dimensions, the stochastic differential equation becomes:
\be
{\rm d}\bx =\bb \, {\rm d}t +{\rm d}\bw,
\label{87}
\ee
where $w_x$, $w_y$, and $w_z$ are assumed to be independent and to have the same diffusion constant $\nu$:
\be
E(w_i (t) w_j (s)) =\nu\delta_{i,j} \min (t,s).
\label{88}
\ee
The forward and backward equations become vector equations.  For $t>s$ one finds
\bea
&&FE: \; \frac{\partial}{\partial t} P(\bx, t ; \by,s) +\nabla_x \cdot \bb(\bx,t) P(\bx, t ; \by,s) -\frac{1}{2} \nu \Delta_x P(\bx, t; \by,s)=0,\label{89}\\
&&BE: \; \frac{\partial}{\partial s} P(\bx, t ; \by, s) +\bb(\by, t)\cdot \nabla_y P (\bx, t; s,\by) +\frac{\nu}{2} \Delta_y P(\bx, t ; \by, s) =0.
\label{90}
\eea
The Hilbert space  $H_t$ can be formed and an operator $\dot{x}$ defined as in the one dimensional case.  The velocity operator satisfies
\be
(f, \dot{\bx}g) = \int {\rm d}^3 x \rho (\bx, t) f (\bx) \left[\bb \left(\bx,t\right)+\nu \bnab\right] g(\bx).
\label{91}
\ee
Likewise, $\ddot{\bx}$ can be defined, and one finds:
\be
\ddot{\bx} =\frac{\partial}{\partial t} \bb +\frac{\nu}{2} \Delta \bb +\frac{1}{2} \nabla \bb^2 +(\bnab\times \bb) \times (\bb +\nu \bnab)
\label{92}
\ee
as an operator equation.

Examining (\ref{92}), one sees that in general $\ddot{\bx}$ cannot be written as a gradient of a potential function $U$.  The acceleration becomes a gradient only if
\be
\bnab\times \bb=0.
\label{93}
\ee
This case shall be considered in this paper.  Magnetic forces require the non-vanishing of (\ref{93}), but these shall be deferred to a future work.

The commutation rules implied by (\ref{91}) are
\be
\left[x_i, x_j\right] =0, \quad \left[\dot{x}_i , x_j\right] =\nu \delta_{i,j}, \quad \left[\dot{x}_i, \dot{x}_j\right] = \nu (\partial_i b_j -\partial_j b_i),
\label{94}
\ee
and with the condition (\ref{93}), the last commutator in (\ref{94}) vanishes, so that the different velocities commute.

If (\ref{93}) is satisfied, then one may write
\bea
R&=& \frac{1}{2} \ln (\rho),\label{95}\\
b&=& \nu \nabla(R+S),
\label{96}
\eea
where (\ref{96}) is the defining equation for $S$ which is defined only up to an arbitrary additive function of $t$.  Eq.\ (\ref{92}) becomes
\be
\ddot{\bx} =\nabla \left[ \nu \frac{\partial}{\partial t} \left(R+S\right) +\frac{1}{2} \nu^2 \Delta \left(R+S\right) +\frac{1}{2} \nu^2 \left(\nabla \left(R+S\right) \right)^2\right] =-\nabla U
\label{97}
\ee
and this equation also defines the potential $U$.

Analogous to eq.\ (\ref{76}), $\rho$ must satisfy
\be
\frac{\partial \rho}{\partial t} +\nabla \cdot \bb\rho -\frac{1}{2} \nu \Delta \rho=0,
\label{98}
\ee
but (\ref{97}) and (\ref{98}) are equivalent to the Markov wave equations:
\be
\left[ \frac{\nu^2}{2}\Delta +U\right] e^{R\pm S} =\mp \nu \frac{\partial}{\partial t} \, e^{R\pm S}.
\label{99}
\ee
Either (\ref{98}) or (\ref{99}) may be used to derive the equation of continuity
\be
\frac{\partial \rho}{\partial t} =-\nabla \cdot \nu \rho \nabla S,
\label{100}
\ee
so that the probability flux may be identified with $\nu \rho \nabla S$, and the velocity field of the diffusion with $\nu \nabla S$.

In the same sense as in section 3, time derivatives of operators may be calculated
\be
\frac{\rm d}{{\rm d}t} f(\bx, \dot{\bx}) =\frac{1}{\nu} \left[ H, f\right] ,
\label{101}
\ee
where
\be
H =\frac{1}{2} \dot{\bx}^2 +U.
\label{102}
\ee

Some useful formulae which may be derived are
\bea
\lim\limits_{\epsilon \to 0} E(\dot{\bx} (t)) &=& \int {\rm d}^3 x \rho (\bx, t) \nu \nabla S,\label{103}\\
\lim\limits_{\epsilon\to 0} \left\{ E((\dot{\bx})^2 ) -\frac{3\nu}{\epsilon}\right\} &=&(1, \dot{\bx}^2)= \int \,{\rm d}^3 x \rho (\bx, t) \nu^2 \left[ \left(\nabla S\right)^2 -\left( \nabla R\right)^2 \right],\label{104}\\
(1, f (\bx, \dot{\bx})) &=& \int {\rm d}^3 x e^{R-S} f\left(\bx, \nu \nabla \right) e^{R+S}.
\label{105}
\eea

The heuristic arguments of section 3 could equally well be applied here, but since they are the same, they will not be repeated.  The restriction $\nabla \times \bb=0$ is essentially the same as Nelson's $\nabla \times \bv =0$ \cite{nelson2}.

\section{Dynamics}

In this section the Markov wave equations (\ref{99}) are considered.  They shall be used to model a diffusing particle of mass $m$.  Multiplying (\ref{99}) by $m$ yields
\be
\left[ \frac{1}{2} m \nu^2 \Delta +\widetilde V\right] e^{R\pm S} = \mp m \nu \, e^{R\pm S} , \qquad \widetilde V =m U,
\label{106}
\ee
where $\widetilde V$ has units of energy.  Let the diffusing particle be subject to an external potential $V$.  Equating $\widetilde V$ and $V$ does not work.  For example if $V=1/2 kx^2$, a harmonic oscillator, then one expects a steady state solution with $\rho=e^{2R}$ and $S=-\lambda t$,  $\lambda$  a constant.  The equations become in this case
\bdm
\left[\frac{1}{2} m\nu^2 \Delta +\frac{1}{2} kx^2\right] e^R =\lambda m \nu \, e^R.
\edm
The solutions to this equation are not normalizable and are unacceptable on physical grounds since they do not even vanish at infinity.

Nelson's dynamical assumption \cite{nelson1,nelson2} equated the mean acceleration to the external force, and this led to Schr\"odinger's equation.  His dynamics are a special case of a more general dynamical system which shall now be presented.

Since $\widetilde V$ and $V$ cannot be equated, it shall be assumed that
\be
\widetilde V=V+h,
\label{107}
\ee
where $h$ is some function of $R$ and $S$.  Translation, rotation, and Galilean invariance shall be assumed in the limit $V\to 0$.  Symmetry breaking effects such as viscous damping can be included later.  With these symmetires  $h$ must yeild no net force or torque:
\be
\int \;{\rm d}^3 x \rho (x, t) (-\nabla h) =\int \; {\rm d}^3 x \rho (x,t) \left[ x_i \partial_j -x_j \partial_i\right] h=0.
\label{108}
\ee
In addition to these constraints, the dynamics of the diffusing particle shall be assumed not to depend on the overall normalization of $\rho$.  That is if $\rho$ is a solution, then $\lambda \rho$ should also be for $\lambda =$ constant.  The reason for demanding this is that if $\rho$ is non vanishing in two widely separated regions, but zero elsewhere, then the diffusion in the two regions should be independent.  This is possible only if the above condition is satisfied.

Two terms have been found for $h$ which satisfy all these conditions.  It is likely that they are not unique, but they do lead to an interesting theory.  The form for consideration is
\be
 h=cR -\lambda m \nu^2 \, \frac{\Delta \sqrt\rho}{\sqrt\rho} , \quad \lambda \mbox{ and } c \mbox{ constants}.
\label{109}
\ee
The reader may verify that all of the conditions are satisfied.

A viscous damping force can be included by adding a term $\kappa S$, $\kappa$ a positive constant, to $V$.  The final form for the equations are:
\be
\left[ \frac{1}{2}m \nu^2 \Delta +V (x) -\lambda m \nu^2 \frac{\Delta \sqrt \rho}{\sqrt \rho}\, +cR +\kappa S\right] e^{R\pm S} =\mp m \nu \frac{\partial}{\partial t} e^{R\pm S}.
\label{110}
\ee
These equations have a remarkable property.  Defining
\be
\beta =\sqrt{\frac{1}{1-2\lambda}},
\label{111}
\ee
then they are equivalent to
\be
\left[\left( \frac{1}{2} -\lambda\right) m\nu^2 \Delta +cR +\kappa S +V \right] e^{R\pm \beta S} =\mp \frac{m\nu}{\beta} \,\frac{\partial}{\partial t} \,e^{R\pm \beta  S},
\label{112}
\ee
so long as $\lambda \not= \frac{1}{2}$, in which case $\beta =\infty$.

The case $\lambda >\frac{1}{2}$ is of particular interest.  It yields an imaginary number for $\beta$:
\be
\beta =i |\beta|,\qquad \lambda >\frac{1}{2}.
\label{113}
\ee
Defining a complex ``wave function''
\be
\psi (x) =e^{R+i|\beta| S},
\label{114}
\ee
 the equations become
\be
\left[ -\frac{1}{2} m \frac{\nu^2}{|\beta|^2} \,\Delta +cR +\kappa S +V\right] \psi =i \frac{m\nu}{|\beta|} \, \frac{\partial \psi}{\partial t}
\label{115}
\ee
together with the complex conjugate of (\ref{115}).  This equation is similar to quantum mechanics.  In fact, for the special case $c=0$, $\kappa =0$, with
\be
\hbar =\frac{m\nu}{|\beta|} =m\nu \sqrt{2\lambda -1},
\label{116}
\ee
then (\ref{115}) becomes
\be
\left[ -\frac{1}{2} \,\frac{\hbar^2}{m} \Delta +V\right] \psi =i\hbar \, \frac{\partial \psi}{\partial t},
\label{117}
\ee
which is Schr\"odinger's equation.    Nelson \cite{nelson2} has already pointed out this connection.  His dynamical  assumption is equivalent to choosing $|\beta|=1$, or $\lambda =1$. It is seen from (\ref{116}) that this is only one possibility.  Any three numbers $m$, $\nu$, and $\lambda$ which satisfy (\ref{116}) will yield Schr\"odinger's equation.  This means there is an infinity of Markov processes, each with different $\nu$, which lead to Schr\"odinger's equation.  One can even consider  ``fictitious'' processes for which $\nu$ is imaginary which lead to the same equation, but which do not strictly speaking have a physical interpretation other than statistical.  It may be that the traditional interpretation of quantum mechanics is based on just such a fictitious picture of a real Markov process.

A second application of the case $\lambda >\frac{1}{2}$ concerns the quantum-mechanical Gibbs distribution.  Consider a particle weakly interacting with a thermostat at temperature $T$.  Let this particle be subject to an external potential well $V(x)$.  The Gibbs distribution for this particle is
\be
\rho_G (x) =\sum\limits^\infty_{n=1}\, \psi_n (x) e^{-E_n /T} \, \psi_n (x).
\label{118}
\ee
This expression has an expansion in terms of  $\hbar$. Landau \cite{landau} (p.\ 99) finds the following form for $R$ up to fourth order corrections in $\hbar$, and up to a normalization constant
\be
R(x) =\frac{1}{2} \ln (\rho_G)=-\frac{1}{2T} \left\{ V-\frac{\hbar^2}{24T^2} \, \frac{1}{m} \,\left( \nabla V\right)^2 +\frac{\hbar^2}{12T}\, \frac{1}{m}\, \Delta V\right\}.
\label{119}
\ee
Compare this expression with (\ref{115}).  Set $S$ to zero since the Gibbs distribution is a steady state.  Consider the following choice of variables:
\be
\frac{m\nu}{ |\beta|} =\frac{\hbar}{\sqrt 3}, \qquad C=2T.
\label{120}
\ee
Eq.\ (\ref{115}) becomes
\be
\left[ -\frac{\hbar^2}{6m} \Delta +2TR +V\right] e^R=0.
\label{121}
\ee
This equation agrees with (\ref{119}) up to fourth order corrections in $\hbar$.  To see this, rewrite (\ref{121}) as
\be
R=-\frac{1}{2T} \left[ V-\frac{\hbar^2}{6m} \, \left( \Delta R+\left( \nabla R\right)^2\right) \right].
\label{122}
\ee
Now iterate
\bea
R_1&=& -\frac{V}{2T},\label{123}\\
R_2 &=& -\frac{1}{2T} \left[ V-\frac{\hbar^2}{6m} \left( \Delta  R_1 +\left( \nabla R_1\right)^2\right) \right]\nn\\
&=& -\frac{1}{2T} \left[ V+ \frac{\hbar^2}{12mT}\, \Delta V -\frac{\hbar^2}{24 mT^2} \left( \nabla V\right)^2\right],
\label{124}
\eea
which agrees with Landau's result (\ref{119}).  This is an encouraging result.  Further iteration does not lead to the correct fourth order term and so the model is only good for small $\hbar$.

A time dependent model is possible within the framework of the model.  Using (\ref{120}), it is
\bea
&& \left[ -\frac{\hbar^2}{6m} \Delta +2TR +V +\kappa S\right] \psi =i \frac{\hbar}{\sqrt 3} \, \frac{\partial \psi}{\partial t},\label{125}\\
&& \psi =\exp \left( R+i \,\frac{\sqrt 3 m\nu }{\hbar} \, S\right) .
\label{126}
\eea
This equation is more useful if one defines
\be
\widetilde S =\nu S
\label{127}
\ee
so that the flow velocity is $\nabla \widetilde  S$.  Then (\ref{125}) becomes
\bea
&&\left[ -\frac{\hbar^2}{6m} \Delta +2TR +V +\frac{\kappa}{\nu} \widetilde S\right] \exp \left( R\pm i \frac{\sqrt 3m}{\hbar} \, \widetilde S\right)\nn\\
&&\,\quad =\pm i \frac{\hbar}{\sqrt 3}  \, \frac{\partial}{\partial t} \exp \left( R\pm i \frac{\sqrt{3}m}{\hbar} \, \widetilde S\right).
\label{128}
\eea
The parameter $\nu/\kappa$ is called the mobility $\mu$ and is easily measured.  In steady state flow, (\ref{128}) becomes
\bea
&&\frac{1}{2} m \left( \nabla \widetilde S\right)^2 -\frac{\hbar^2}{6m} \,\frac{\Delta \sqrt \rho}{\sqrt \rho} +2TR +V +\frac{1}{\mu} \widetilde S =\mbox{ constant},\label{129}\\
&&\nabla \cdot \rho \nabla \widetilde S =0.
\label{130}
\eea
At high $T$ and slow flow, (\ref{129}) becomes
\be
2TR +V +\frac{1}{\mu} \,\widetilde S =\mbox{ constant},
\label{131}
\ee
which yield upon taking a gradient
\be
\mbox{Flux } =\rho \nabla \widetilde S =-\mu T\nabla \rho - \mu \nabla V.
\label{132}
\ee
In most practical applications, eq.\ (\ref{132}) for the probability flux is used together with the equation of continuity
\bdm
\frac{\partial \rho}{\partial t} =- \nabla \cdot \rho \nabla \widetilde S.
\edm
The two together yield
\be
\frac{\partial \rho}{\partial t} =D \Delta \rho +\mu \Delta V,  \qquad D=\mu T,
\label{133}
\ee
which is the usual diffusion equation used in physics.  It is only approximate, and in the Markov description it is a kind of quasi-static approximation.  When flow velocities are high, or when quantum effects are important, then (\ref{133}) is invalid.

A Newtonian limit is obtained from the Markov equations (\ref{128}) when the diffusing particle is well localized.  The following may be derived by direct computation using (\ref{128}),
\be
m\frac{\partial^2}{\partial t^2} \, E(x(t)) =\int \, {\rm d}^3 x \rho (x,t) (-\nabla V)-\frac{1}{\mu} \, \frac{\partial}{\partial t} \, E (x(t)).
\label{134}
\ee
If $\nabla V$ changes slowly over the region where $\rho$ is non vanishing, then if $x(t)$ is the mean position of particle, one obtains
\be
m \ddot{\bx} \approx -\nabla V(x) -\frac{1}{\mu} \dot{\bx},
\label{135}
\ee
as a trajectory equation which is of course just Newton's equation.

In the limit $\hbar \to 0$, eq.\ (\ref{128}) becomes equivalent to two equations:
\bea
&& \frac{1}{2} m \left( \nabla \widetilde S\right)^2 +2TR +V +\frac{1}{\mu} \widetilde S =-m\dts,
\label{136}\\
&& \frac{\partial \rho}{\partial t} =-\nabla\cdot \rho \nabla \widetilde S.
\label{137}
\eea
To understand these, consider ignoring the second and fourth terms in (\ref{136}). It becomes
\be
\frac{1}{2} m \left( \nabla \widetilde S \right)^2+ V =-m\dts.
\label{138}
\ee
This is essentially the Hamilton-Jacobi equation for classical mechanics.  Initial conditions on $R$ and $S$ must be specified for a solution.

As an example, consider the case
\be
\widetilde S =\bVV\cdot \bx, \quad t=0,\quad V= \mbox{ constant}.
\label{139}
\ee
With this initial condition, eq.\ (\ref{138}) can be solved for $\widetilde S$.  Define $x(t, x_0)$ by
\be
\frac{\partial}{\partial t} \bx(t, \bx_0)=\nabla \widetilde S (\bx (t, \bx_0) , t), \qquad \bx(0, \bx_0) =\bx_0.
\label{140}
\ee
Then $x(t, x_0)$ is a solution to Newton's equation
\be
\frac{\partial^2}{\partial t^2} \, m\bx (t, \bx_0) =-\nabla \bVV (\bx (t, \bx_0)), \quad \bx(0, \bx_0) =\bx_0, \quad \dot{\bx} (0, \bx_0)=\bVV.
\label{141}
\ee
The solution to (\ref{137}) for $\rho$ is
\be
\rho (\bx,t) =\int\, {\rm d}^3 x_0 \rho (\bx_0, 0) \delta^3 (\bx-\bx(t, \bx_0)).
\label{142}
\ee
This describes the statistical evolution of an ensemble of Newtonian particles all with initial velocity $V$, but having different starting positions distributed by $\rho (x_0, 0)$.

The viscous term in (\ref{136}) can also be included in this manner:
\be
-m\dts =\frac{1}{2} m \left( \nabla \widetilde S\right)^2 +\frac{1}{\mu} \widetilde S +V (x).
\label{143}
\ee
Defining $x(t, x_0)$ by (\ref{140}) one finds
\be
m\frac{\partial^2 x}{\partial t^2} =-\nabla V -\frac{1}{\mu} \,\dot{x}, \quad x(0, x_0) =x_0, \quad \dot{x} (0, x_0) =V,
\label{144}
\ee
and $\rho$ is again obtained from (\ref{142}).  The interpretation is the same except that the ensemble of Newtonian particles is subject to an additional viscous force.

The only outstanding term in eq.\ (\ref{136}) in this Hamilton-Jacobi point of view is the $2TR$ term.  It can be included iteratively.  Consider again the initial condition $\widetilde S=V\cdot x$ as an example.  First ignore $2TR$, solve (\ref{134}) and (\ref{144}), and calculate $\rho$ from (\ref{142}).  The first approximation to $R$ is
\be
R_1 =\frac{1}{2} \ln \left[ \int {\rm d}^3 x_0 \rho (x_0, 0) \delta^3 \left(x-x\left(t, x_0\right)\right)\right].
\label{145}
\ee
This function can next be substituted for $R$ in eq.\ (\ref{136}).  Qualitatively this term acts like  a repulsive potential forcing the particles away from the center of mass.  This causes the distribution to spread more than the spreading inherent in the Newtonian ensemble.  A similar treatment is possible for the general model (\ref{115}), but his will not be presented here.

The $\hbar \to 0$ limit is equivalent to the limit $\lambda =\frac{1}{2}$.  As a final point, the case $\lambda <\frac{1}{2}$ in (\ref{112}) leads to unphysical solutions as can be seen by examining the steady state harmonic oscillator.

The model which has been presented offers a possible description of diffusion when quantum effects cannot be ignored, and to second order accuracy in $\hbar$.  In its present form it cannot be made to agree to all orders of $\hbar$ with the Gibbs distribution in the steady state limit.  Undoubtedly part of the problem is that $h$ in (\ref{109}) is too simple a function in the present formulation.  The only other work in which the author is aware of which attempts a description of joint thermal and quantum diffusion is the work of Schwinger \cite{schwinger}, but the connection with the present theory is not clear at the present time.

\section{Conclusion}

The non-commutative approach presented here is a natural way to analyze Markov processes and diffusion.  It dramatically suggests a deep connection between the Markov theory and quantum mechanics.  The fact that the Markov description of Schr\"odinger's equation is not unique is a new and presumably relevant result for the stochastic interpretation of quantum mechanics.  The models presented for diffusion with quantum effects, if verified, may have practical value.  The $\hbar\to 0$ limit of these may, under certain circumstances, be an alternative to the Ornstein-Uhlenbeck model of Brownian motion.

\bigskip

\bigskip

\end{document}